\documentclass[11pt]{iopart}

\usepackage{iopams}
\usepackage{graphicx}   
\usepackage{ulem}
\usepackage{cancel}
\usepackage{bm}         
\usepackage{float}
\usepackage{tikz}       
\usepackage{color,soul} 
\usepackage{subcaption}
\newcommand{\tikzcircle}[2][red,fill=red]{\tikz[baseline=-0.5ex]\draw[#1,radius=#2] (0,0) circle ;}%

\begin{document}

\title[]{On the existence of an intermediate phase in the antiferromagnetic Ising model on the face-centered cubic lattice}

\author{Graeme Ackland$^{\dagger}$}
\address{School of Physics and Astronomy, University of Edinburgh, Edinburgh EH9 3FD, United Kingdom}
\ead{gjackland@ed.ac.uk}
\vspace{10pt}
\begin{indented}
\item[] \date{\today}
\end{indented}

\graphicspath{{figs/}}

\begin{abstract}
  We use Monte Carlo simulation to determine the stable structures in the second-neighbour Ising model on the face-centred cubic lattice. Those structures are L1$_1$ for strongly antiferromagnetic second neighbour interactions and L1$_0$ for ferromagnetic and weakly antiferromagnetic second neighbours. We find a third stable "intermediate" antiferromagnetic phase with I4$_1$/amd symmetry, and calculate the paramagnetic transition temperature for each.
  The transition temperature depends strongly on second neighbour interactions which are not frustrated.
Our results contradict a recent paper\cite{jurvcivsinova2022prediction}, which also reported two different AFM structures and a new "intermediate" phase exists in this system.  Here we show that the assumed sublattice structure in \cite{jurvcivsinova2022prediction} is inconsistent with the ground state.  We determine a sublattice structure suitable for solving this problem with mean field theory.   
\end{abstract}

%
\vspace{2pc}


%
%

\section{Introduction} \label{sec:intro}

Calculation of phase stability in the antiferromagnetic Ising model is challenging because of the existence of many possible antiferromagnetic arrangements.   Furthermore, the face-centred cubic lattice (fcc, A1 in Strukturbericht designation), which features triangles of neighbouring atoms, suffers from frustration.  The two main approaches to the problem are Monte Carlo simulation and mean field theories\cite{binder1980ordering,beath2005fcc,beath2006phase,beath2007latent,polgreen1984monte,de2008phase,phu2009crossover}.  Monte Carlo correctly includes all correlation effects, but being a numerical method cannot determine the phase boundary analytically\cite{ackland2006magnetically,ehteshami2021high}.  By contrast, effective mean field approaches\cite{ehteshami2020phase} are typically built on cluster approaches which limits the spatial range of correlations. 

In the language of a magnetic system, the Hamiltonian, $\mathcal{H}$, for the Ising model with the nearest-neighbour (NN) interaction, $J_1$, and the next-nearest-neighbour (NNN) interaction, $J_2$, is

\begin{equation} \label{eq:H}
\mathcal{H} = -J_1 \sum\limits_{\left\langle i,j \right\rangle^{\prime}} S_i S_j -J_2 \sum\limits_{\left\langle i,j \right\rangle^{\prime\prime}} S_i S_j - H \sum\limits_{i=1} S_i, \label{eq:Ising}
\end{equation}
where $\left\langle \right\rangle^{\prime}$ stands for summation over NNs, and $\left\langle \right\rangle^{\prime\prime}$ for NNNs. Ising spins $S_i$ are taken as $\pm 1$.
 $H$ is the magnetic field which we consider only in the ground state analysis; simulations are at zero field ($H=0$).
The Hamiltonian in the above equation \ref{eq:Ising} can be analysed as a function of two dimensionless quantities: the ratio of the interactions relative to each other, and to the temperature.
\begin{equation}
\hspace{2cm} \alpha = J_2/|J_1|, \hspace{1cm} \beta^{-1} = T/|J_1|.
\end{equation}

Without loss of generality, we choose units such that $|J_1|=1$.

Many previous authors have looked at the near-neighbour only case\cite{mackenzie1981low,finel1986phase,gahn1986ordering,mazel1988ising,zarkevich2008low,lundow2009ising,stubel2018finite} In our previous work\cite{ehteshami2020phase}, we analysed the case where $\alpha$ is positive, i.e. second neighbour interactions are ferromagnetic.  We also considered non-zero field, creating a three-dimensional $\alpha,$ $T,$ $H$ phase diagram.   In that system the possible phases are L1$_0$, L1$_2$ and paramagnetic.  Those phases were examined in mean field theory using a conventional (4-atom) fcc cell in which the four sites are treated the independent sublattices.   A superdegenerate point exists at H=4, T=0 where  L1$_0$, and L1$_2$ are degenerate, as are a range of point and extended defects.

Recently, Jurčišinová and  Jurčišin (JJ)\cite{jurvcivsinova2022prediction} tackled the harder problem of  $\alpha<0$, where second neighbour interactions are also antiferromagnetic, simplifying matters by setting $H=0$.   Crucial to this is the choice of sublattice structure.  They used a three-site sublattice structure in which 75\% of sites are type "C" (see Appendix). 
As a consequence, all their reported paramagnetic structure have a finite magnetisation. They reported that the phase diagram has two "antiferromagnetic" phases (named AFM1 and AFM2) and a third "well-defined" intermediate phase.
Here we investigate whether the spontaneously-magnetized structures reported by JJ\cite{jurvcivsinova2022prediction} are stable, first by analytic means at zero temperature, then numerically at finite temperatures.  For completeness, we consider both ferromagnetic and antiferromagnetic $J_1$.

\section{Ground State structures}

\begin{table}[t!]
	\begin{center}
		\begin{tabular}{|l|ccc|}
			\hline
 Structure  &Free energy & 	Magnetization & Stability
   \\ \hline
			 L1$_0$ & $-4J_1+6J_2$ &0 & AFM $J_1$, FM $J_2$ \\   
   I4$_1$/amd      & $-4J_1+2J_2$ & 0  &  $J_1$, AFM $J_2$,  \\ 
                L1$_1$ & $-6J_2$ & 0  &   AFM $J_2$, $J_1<-J_2$ \\ 
    Ferromagnetic & $12J_1+6J_2-H$ & 1 & FM $J_1$, FM $J_2$ \\
    Paramagnetic & 0 &0 & high T \\
    \hline
    Ferromagnetic\cite{ehteshami2020phase} & $12J_1+6J_2-H$ & 1 & high H \\
    DO$_{22}$ \cite{ehteshami2020phase}& $2J_2-H/2$ & 1/2 & AFM $J_1$, AFM $J_2$, medium H\\
    AFM1\cite{jurvcivsinova2022prediction} (L1$_2$ & $6J_2-H/2$ & 1/2 & AFM $J_1$, FM $J_2$, medium H \\
    AFM2\cite{jurvcivsinova2022prediction} ($m_C$=1) & $12J_1+9J_2/2 -3H/4
    $   & 3/4 & nowhere \\
    AFM2\cite{jurvcivsinova2022prediction} ($m_C$=0) & -1.5$J_2$
       & 0 & nowhere \\
   \hline    
		\end{tabular}
	\end{center}
	\caption{Perfect crystal energies at T=0.  AFM1 and AFM2 are from Ref \cite{jurvcivsinova2022prediction}.  "Stability" indicates the region of the phase diagram where the phase is expected. Horizontal line separates phases observed in this work from others reported elsewhere. \label{tab:tableY}}
\end{table}

\begin{figure}[tp] 
	\centering
		\includegraphics[width=0.33\linewidth]{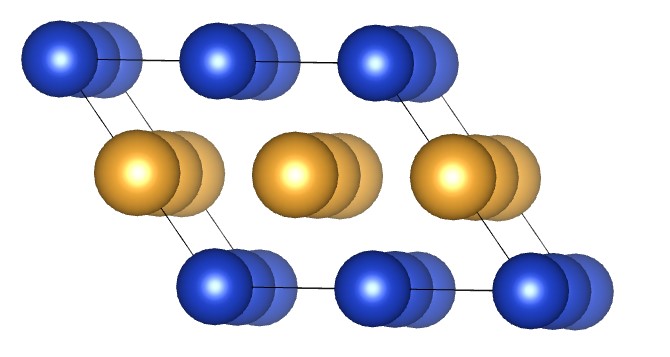}%
  	\includegraphics[width=0.33\linewidth]{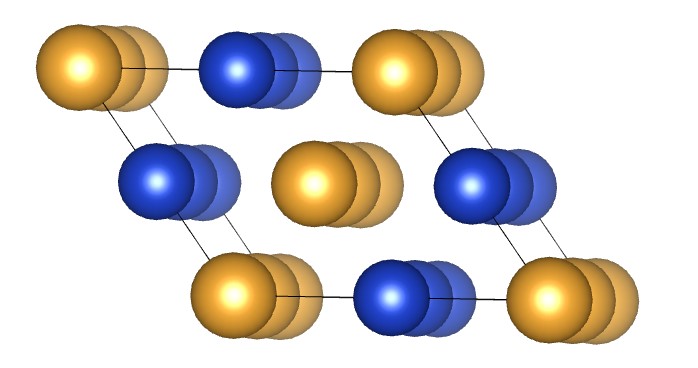}%
    	\includegraphics[width=0.33\linewidth]{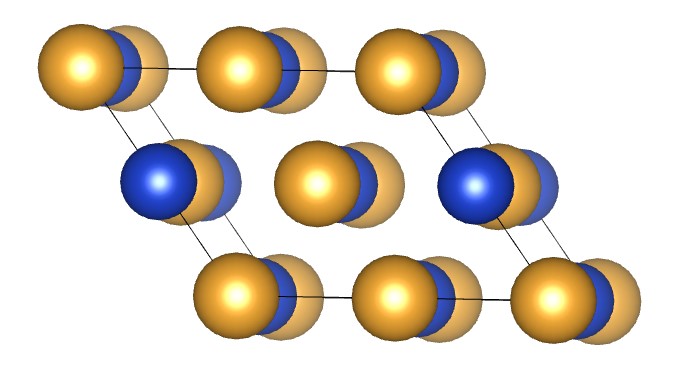}%
	\caption{The FCC lattice in the $a=(110), b=(1,\bar{1},0), c=(\frac{1}{2},\frac{1}{2},1)$ setting viewed close to the (110) direction.   Colouring shows the patterns of the various sublattice spin ordering corresponding to the L1$_0$, L1$_1$ and I4$_1$/amd structures.}
\label{fig:sub44}\end{figure}

First we consider only the T=0 case, attempting to identify the possible stable structures. According to the Third Law of thermodynamics, an ordered state must be the most stable. Identifying these candidate states is a necessary precursor to making a sensible definition of order parameters or sublattice structures. The relevant phases are shown in Figure \ref{fig:sub44} with details given in Table \ref{tab:tableY} and the Appendix.

If we consider the ground state of the JJ structures, we see that AF1 has $m_A=m_B=-m_C$.  This is the L1$_2$ structure, which can be obtained in the four-sublattice model with $m_1=m_2=m_3=-m_4$, with a ground state energy being a weighted average: 

\begin{eqnarray*}  E_{L1_2}& = &E_A/8+E_B/8+3E_C/4  \\ 
&=& 0.125(12J_1-6J_2) + 0.125(12J_1-6J_2) +  0.75(-4J_1-6J_2)\\
&=& -6J_2.
\end{eqnarray*}
For antiferromagnetic $J_2$ this is less stable than randomly oriented spins, and therefore L1$_2$ (AF1) should not  appear in this region of the phase diagram, since it is not stable at T=0, and has lower entropy than the disordered paramagnetic state. DO$_{22}$ is always more stable than L1$_2$, but even it may only be stabilised by an external field\cite{ehteshami2020phase}.

We can contrast this with the L1$_0$ phase which comprises alternating (001) planes of different spins;  
using our sublattice structure it is $m_1=m_2=-m_3=-m_4$, but L1$_0$ cannot be represented within the  three-sublattice assumption. 
In L1$_0$ all sites have equal energy $E=-4J_1+6J_2$.  This is the unique stable state at zero field for ferromagnetic $J_2$, and extends some way into the antiferromagnetic $J_2$ region (Figure \ref{fig:PD}.  Clearly, for $6J_2>4J_1$ this L1$_0$  structure has higher than zero, so some other ordered phase must exist which favours unlike second neighbours.   

This phase is  L1$_1$  a layered structure with alternating (111) close-packed planes of opposite spins, symmetry $R\overline{3}m$.  It cannot be defined based on either of the sublattices considered above.  Relative to the conventional fcc cell it is a two atom cell with a=(1/2,-1/2,0), b=(-1/2,0,1/2), c=(0,1,-1), with basis atoms at (0,0,0) and (0,0,1/2) which define the sublattice.   This structure has T=0 energy -6J$_2$, and so becomes degenerate with L1$_0$ at $J_2=J_1/3$.   

It seemed unlikely that L1$_0$, which has all NNN aligned, could persist when $J_2$ is antiferromagnetic. For near-neighbour only interactions L1$_0$ has zero-energy stacking faults\cite{ehteshami2020phase}, and by considering an array of stacking faults we found an intermediate phase 
with I4$_1$/amd symmetry which does not appear in the Strukturbericht designation.
This is degenerate with L1$_1$ at $J_2=J_1/2$ and L1$_0$ $J_2=0$, and more stable between those values. 

We note that in the limit $J_1\rightarrow 0$ the fcc structure breaks into four unconnected simple cubic lattices, which can be made independently antiferromagnetic in the B1 (NaCl) structure without frustration.   L1$_1$ can be viewed as four interpenetrating NaCl lattices.

\section{Numerical simulations}
We ran Metropolis Monte Carlo\cite{metropolis1953equation} simulations on a 12x12x12x4 atom supercell.
The model parameters are $J_2$ and $T$ and there are two cases: ferromagnetic $J_1=1$ and antiferromagnetic $J_1=-1$.  No external field was applied $(H=0)$.  Updates were single-site flips, of randomly-chosen sites. At each temperature we equilibrate for 10$^6$ attempted flips and  collect data for 10$^9$.

\begin{figure}[tp] 
	\centering
		\includegraphics[width=0.5\linewidth]{
  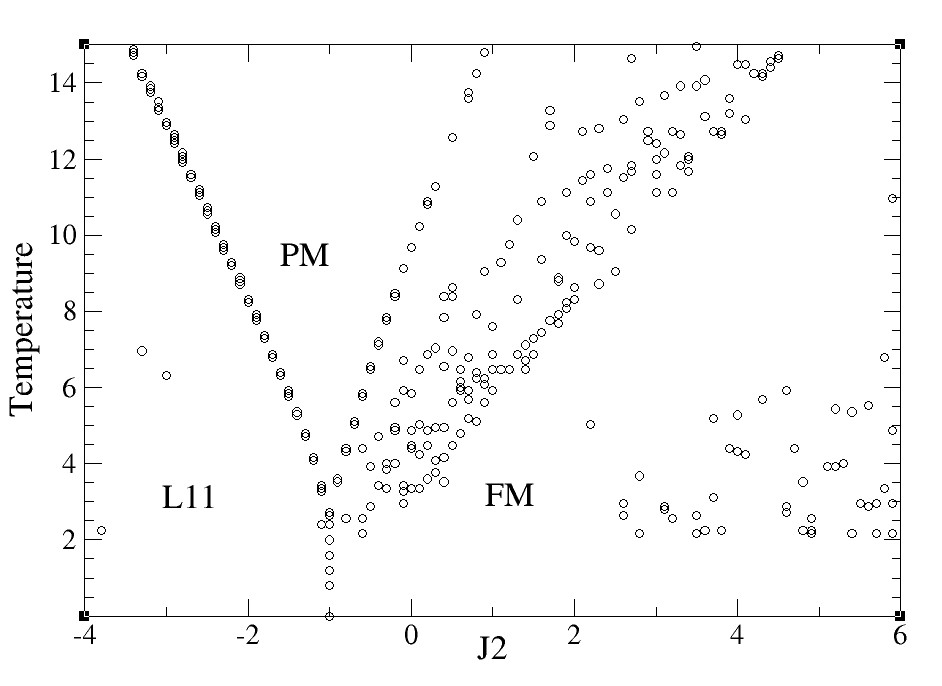}%
  	\includegraphics[width=0.5\linewidth]{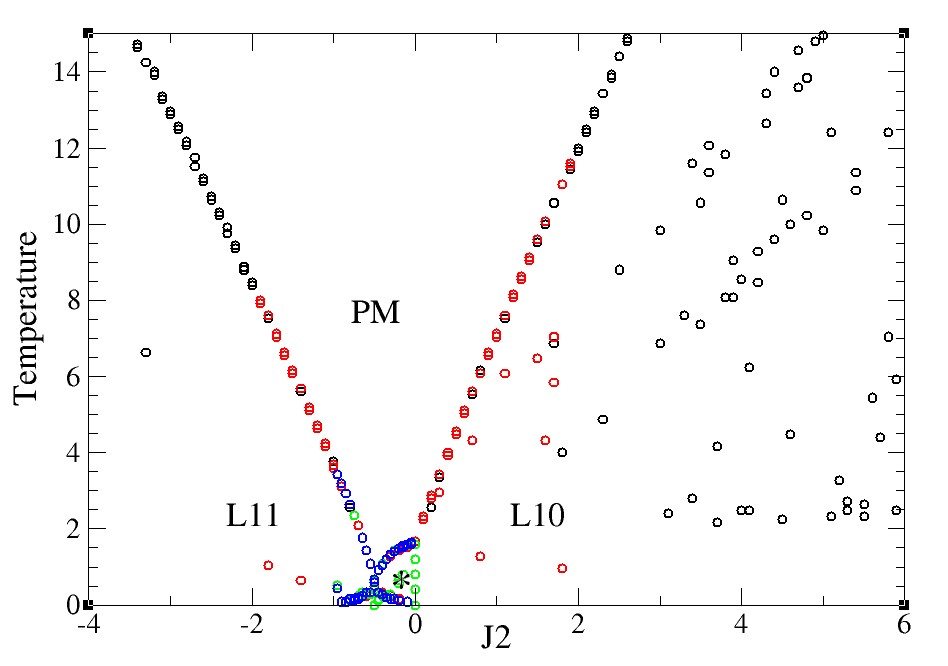}%
	\caption{ Phase diagram for (left) Ferromagnetic $J_1=1$ (right) Antiferromagnetic $J_1=-1$.   Points indicate the ($J_2,T$)  tuple for the  two highest values of peaks in $c$: for the PM transition line this is a lambda peak, within ordered phase is comes from annealing a domain structure.  Colours indicate starting configuration: black: PM,  red: FM, blue L1$_0$, green L1$_1$. Star indicates the small region of I4$_1$/amd.     } 
\label{fig:PD}\end{figure}

In Figure \ref{fig:PD} we show the phase diagram found by monitoring the temperature variation of fluctuations in the energy:
\begin{equation}
c(T)=<\mathcal{H}^2> - <\mathcal{H}>^2
\end{equation}
and detecting peaks therein.  To detect transitions between ordered phases we monitor fluctuations in the NNN contribution to the energy only. 

The simulations revealed just four distinct ordered phases, all of which were as anticipated from the analytic ground state calculations.
\begin{itemize}
\item
ferromagnetic  for $J_1>0; J_2>-J_1$,
\item L$1_0$ for
$J_1<0; J_2>0$, 
\item
I4$_1$/amd for  $J_1<0;$ $-J_1/2<J_2<0$,
\item L1$_1$ for $J_1<0; J_2<-J_1/2$, and for $J_1>0; J_2<-J_1$.
\end{itemize}

The AFM1 and AFM2 structures proposed by JJ are not observed, and if the simulation is initiated in AFM2 it is unstable.
Our intermediate I4$_1$/amd structure is also different from the JJ intermediate structure.

Peak detection is not completely straightforward, because a high variation of 
$\mathcal{H}$ can occur if there is a domain structure which rearranges itself during a simulation.  Such an event produces a high $c(T)$ at a single temperature, whereas a thermodynamic phase transition produces a characteristic lambda transition across a range of temperatures.  To address this, we plot in Fig.\ref{fig:PD} the temperatures corresponding to the two highest values of $c(T)$ as points on a graph of $J_2$ vs $T$.  This traces out the phase boundaries with a sharp line, and also shows a diffuse region corresponding to the "annealing temperature", at which point the single-flip algorithm is able to anneal out a domain structure.   It is notable that the L1$_1$ structure appears less susceptible to domain formation than other phases.

The phase lines are rather straight, with the PM transition temperature lowest at the "maximally frustrated" value of $J_2$ where two ordered structures are degenerate.

\section{Sublattice structures}

A mean field treatment of the antiferromagnetic second neighbour Ising model will require a sublattice decomposition which permits all possible ground states: alternating (001) layers and alternating (111) layers, and the $I4_1/amd$.  Each have two independent sublattices, so a supercell which can describe them all requires at least eight sublattices.  One such structure is shown in Fig.\ref{fig:sub44}.   Compared to the conventional fcc cell it has a=(1,1,0) b=(1,-1,0) c=($\frac{1}{2},\frac{1}{2}$,1).   To include L1$_2$ and DO$_{22}$ structures a still larger set of sublattices is needed, based on a 16 atom cell a=(1,1,0) b=(1,-1,0) c=(0,0,2). (Table \ref{tab:sc})

\begin{table}[t!]
	\begin{center}
		\begin{tabular}{|ccc|cccccc|}
			\hline
  x & y & z & L1$_0$ & L1$_1$ & I4$_1$/amd  & L1$_2$ & DO$_{22}$& FM
   \\ \hline
		0 & 0 & 0 & 1&1&1 &1&1&1 \\
  	1/2 & 0 & 0 & 1&1&-1  &1&1&1 \\
   	1/2 & 1/2 & 0 & 1&-1&1 &1&1&1  \\
    	0 & 1/2  & 0 & 1&-1&-1 &1&1&1  \\	
     1/4 & 1/4 & 1/4 & -1&-1&-1  &-1&-1&1 \\
  	1/4 & 3/4 & 1/4 & -1&1&1  &1&1&1 \\
   	3/4 & 1/4 & 1/4 & -1&-1&1  &1&1&1 \\
3/4 & 3/4  & 1/4 & -1&1&-1  &-1&-1&1 \\	
     0 & 0 & 1/2 & 1&-1&-1  &1&1&1 \\
  	1/2 & 0 & 1/2 & 1&-1&1 &1&1&1  \\
   	1/2 & 1/2 & 1/2 & 1&1&-1 &1&1&1  \\
    	0 & 1/2  & 1/2 & 1&1&1 &1&1&1  \\
     1/4 & 1/4 & 3/4 & -1&1&1 &-1&1&1  \\
  	1/4 & 3/4 & 3/4 & -1&-1&-1  &1&-1&1 \\
   	3/4 & 1/4 & 3/4 & -1&1&-1 &1&-1&1  \\
3/4 & 3/4  & 3/4 & -1&-1&1  &-1&1&1   \\
   \hline    
		\end{tabular}
	\end{center}
	\caption{Fraction positions in tetragonal supercell with $a=b=\sqrt{2}$, $c=2$ relative to conventional fcc cell, and associated ground state spins for structures in the phase diagram. \label{tab:sc}}
\end{table}

\section{Discussion and conclusions}

We find four different ordered phases in the second-neighbour ($J_1,J_2$)  Ising model on the $fcc$ lattice:  Ferromagnetic fcc, and ordered AFM phases I4$_1$/amd, L1$_1$, and L1$_0$.  All of these are stable at zero temperature, and with increased temperature, all transform to a paramagnetic state.  

Numerical simulations show that the stable structures with antiferromagnetic J$_1$ interactions all have zero magnetisation (assuming $H$=0).  Spontaneous magnetisation is observed only for ferromagnetic J$_1$.

These results contradict a recent mean field calculation, which also reported two AFM states and an intermediate structure.
We trace the discrepancy to the fact that the 3-sublattice decomposition assumed in that work does not permit the L1$_0$, I4$_1$/amd  and L1$_1$ groundstates of the antiferromagnetic fcc lattice.  Similarly, the 4-sublattice decomposition  which was used previously\cite{ehteshami2021high} in the ferromagnetic $J_2$ would also be inappropriate for the antiferromagnetic $J_2$ case.

The paramagnetic transition temperature is strongly dependent on $J_2$, taking its lowest value at the point where two competing ordered structures have identical ground-state enthalpy.  This is true regardless of whether T is measured in units of $|J_1|$ or an average interaction weighted by number of neighbours, i.e. $|J_1|+|J_2|/2$.  The disproportionate effect of $J_2$ on the transition temperature follows from the absence of frustration in NNN interactions.

\section*{Acknowledgement}
Funding for this work was provided by ERC grant Hecate.  The author thanks Hossein Ehteshami for bringing this problem to his attention.  For the purpose of open access, the author has applied a Creative Commons Attribution (CC BY) licence to any Author Accepted Manuscript version arising from this submission.

\section{References}
\bibliographystyle{iopart-num}
\bibliography{Potts}

\section{Appendix- previous sublattice decompositions}

\begin{figure}[h] 
	\centering
	\begin{subfigure}{0.30\textwidth}
		\includegraphics[trim={0.5cm 0.5cm 0.5cm 0.4cm},clip,width=\linewidth]{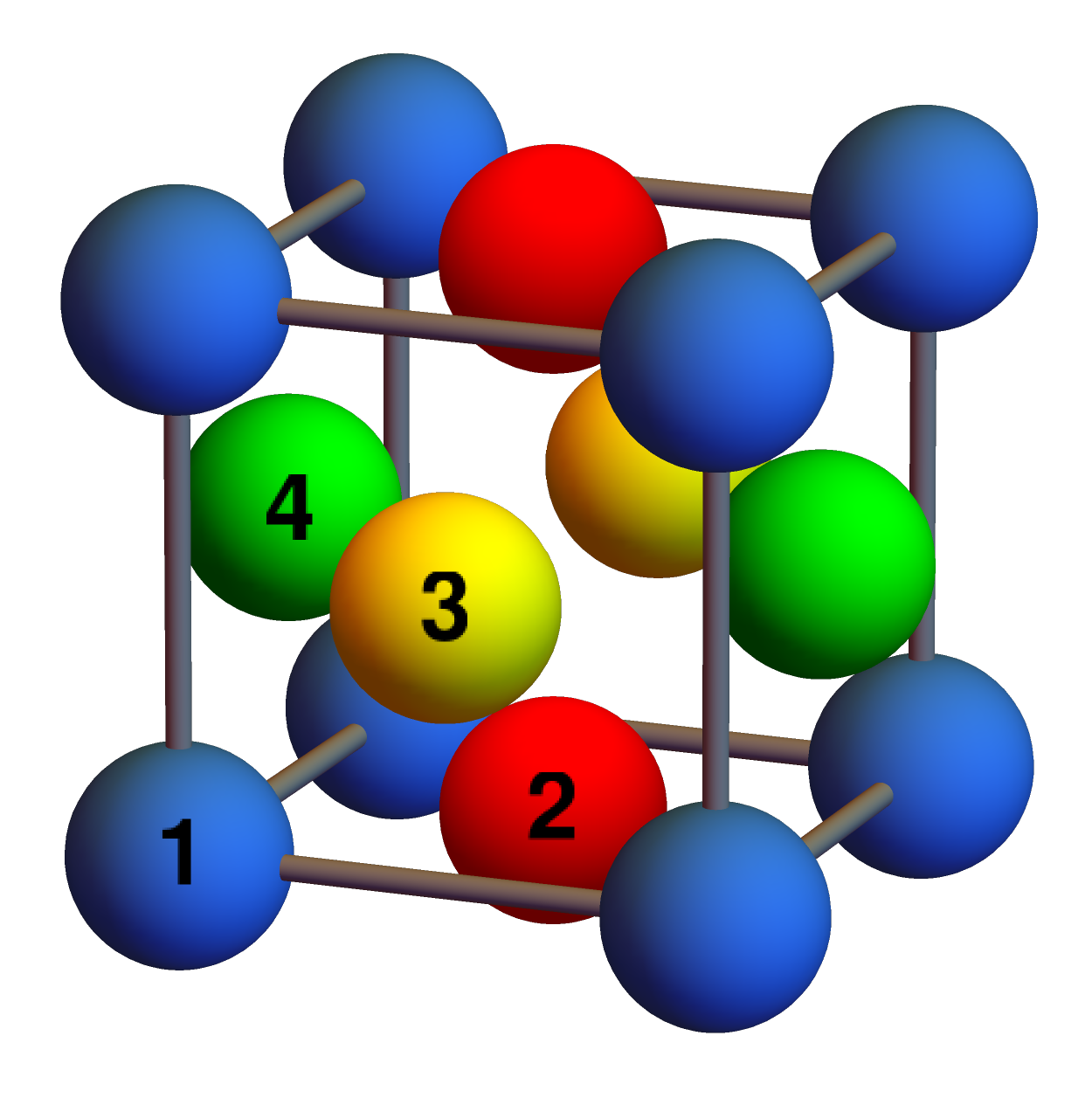}%
		\caption{} \label{fig:fcc}
	\end{subfigure}
	\hspace{1mm}%
	\begin{subfigure}{0.30\textwidth}
		\includegraphics[trim={0.5cm 0.5cm 0.5cm 0.4cm},clip,width=\linewidth]{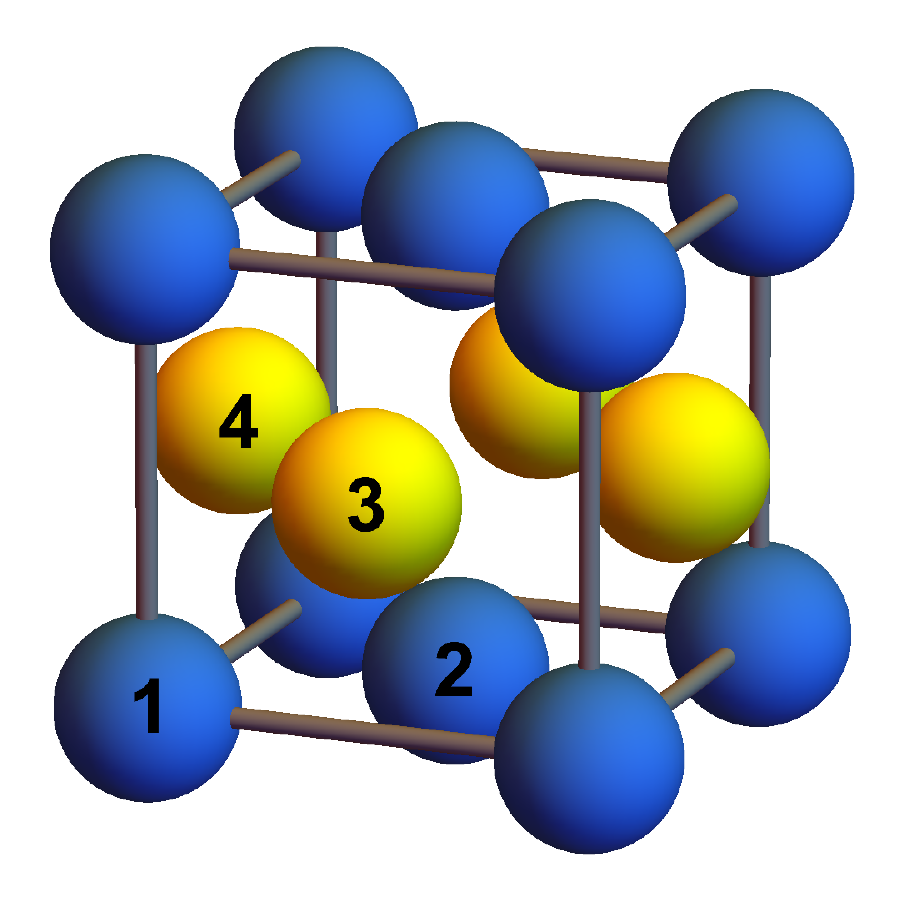}%
		\caption{} \label{fig:L10}
    \end{subfigure}
	\hspace{1mm}%
	\begin{subfigure}{0.30\textwidth}
		\includegraphics[trim={0.5cm 0.5cm 0.5cm 0.4cm},clip,width=\linewidth]{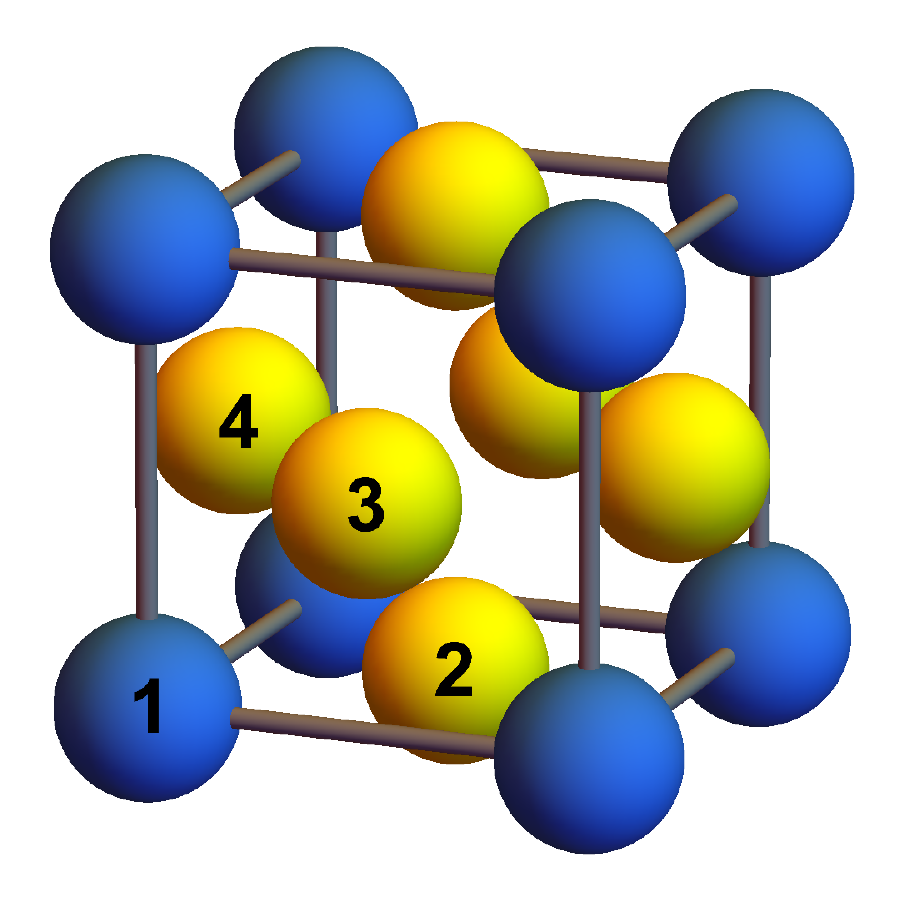}%
		\caption{} \label{fig:L12}
	\end{subfigure}
	\caption{{(a)} Four-sublattice decomposition based on conventional unit-cell of FCC. FCC lattice can be considered as four interpenetrating simple cubic (SC) lattices which each SC lattice here is denoted  by a different color. {(b)} ${\rm L1}_0$ is represented by $A = $ $m_1$ (\tikzcircle[blue,fill=blue]{4pt}) = $m_2$ (\tikzcircle[red,fill=red]{4pt}), $B = $ $m_3$ (\tikzcircle[yellow,fill=yellow]{4pt}) = $m_4$ (\tikzcircle[green,fill=green]{4pt}), and {(c)} ${\rm L1}_2 $ by $A = $ $m_1$ (\tikzcircle[blue,fill=blue]{4pt}), $B = $ $m_3$ (\tikzcircle[yellow,fill=yellow]{4pt}) = $m_2$ (\tikzcircle[red,fill=red]{4pt}) = $m_4$ (\tikzcircle[green,fill=green]{4pt}).}
\end{figure}

\begin{figure}[h] 
	\centering
		\includegraphics[width=0.5\linewidth]{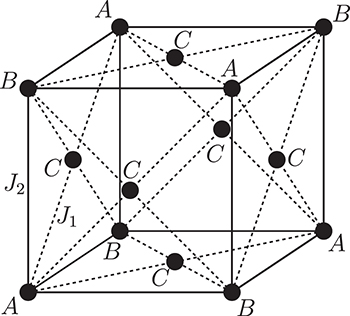}%
	\caption{ Three-sublattice decomposition based on conventional unit-cell of FCC. Figure taken from Jurčišinová and  Jurčišin \cite{jurvcivsinova2022prediction}}
\label{fig:sub3}\end{figure}
\end{document}